\documentclass[amsmat,amssymb,amsfonts,aps,prb,twocolumn,reprint,showpacs]{revtex4-1}

\usepackage{graphicx}
\usepackage{xcolor}

\begin{document}

\title{Electronic properties of  the MoS$_2-$WS$_2$ heterojunction }

\author{K. Ko\'smider$^{1}$, J. Fern\'andez-Rossier$^{1,2}$ }
\affiliation{ ${(1)}$ International Iberian Nanotechnology Laboratory (INL),
Av. Mestre Jos\'e Veiga, 4715-330 Braga, Portugal
\\ ${(2)}$ Departamento de F\'isica Aplicada, Universidad de Alicante, 03690 San Vicente del Raspeig, Spain
}
\date{\today}

\begin{abstract}
We study the electronic structure of  a heterojunction made of two monolayers of  MoS$_2$ and WS$_2$. Our first-principles density functional calculations show that, unlike in the homogeneous bilayers, the  heterojunction has an optically active band-gap, smaller than the ones of  MoS$_2$ and WS$_2$ single layers. We find that that the optically active states of the maximum valence and minimum conduction bands are localized on opposite monolayers, and thus the lowest energy electron-holes pairs are spatially separated. Our findings portrait  the MoS$_2-$WS$_2$ bilayer as a prototypical example for band-gap engineering of atomically thin two-dimensional semiconducting heterostructures.
\end{abstract}

\maketitle

%\pacs{PACS numbers: 71.35.+z}
%\section{Introduction}

Engineering the electronic properties of semiconductors by using heterojunctions has been  the central concept in semiconductor science and technology for 5 decades. \cite{Alferov,Cardona-Yu} With the advent of quantum wells, band-gap engineering of quasi-two dimensional semiconductors made it possible to observe a wealth of new physical phenomena, including the integer and fractional quantum Hall effects in modulation doping GaAs/GaAlAs,\cite{VK,Tsui} the condensation of both excitons in double GaAs quantum wells of GaAs, \cite{Butov.PRL} and exciton-polaritons in II-VI quantum wells \cite{Saba,Deng}  and, more recently, the Quantum Spin Hall phase in CdTe/HgTe quantum wells. \cite{Konig}

The isolation \cite{PNAS}  of truly two dimensional crystals, such as graphene  and  MoS$_2$,  and their use to fabricate field effect transistors, \cite{Novoselov05,Kim05} has opened a wealth of new venues in physics and material science in general, and more specifically  in the design  heterostructures. \cite{Novoselov2012} Thus, graphene bilayers \cite{Novoselov06} and  graphene on boron nitride \cite{Dean}  have different electronic properties than freestanding graphene.

The properties of  bulk MoS$_2$ and its nanostructures, such as nanotubes, \cite{Remskar,Enyashin} fullerenes, \cite{Deepak} and nanoislands, \cite{Helveg-PRL2000} have been studied for a long while, including even chemically exfoliated single planes. \cite{Yang91} More recently,  the study of electronic and optoelectronic devices based on a single MoS$_2$ layer has taken impetus for several reasons. First,  it was found that monolayers of MoS$_2$ feature a direct band gap of 1.8 eV with strong photoluminescence, \cite{Splendiani,Mak_PRL2010} as opposed to bulk MoS$_2$ which has indirect band gap of 1.29 eV. Second, the fabrication of a high mobility field effect transistor based on single MoS$_2$ layer has been reported. \cite{Radisavljevic} Third,  the combination of hexagonal symmetry, large spin-orbit coupling  and lack of inversion symmetry, give rise to  gapped graphene like bands, with two valleys and strong spin-valley coupling. \cite{Xiao} Taking advantage of these unique properties, optical spin pumping is turned into valley-polarized photo carriers, \cite{Cao,ZengNature,Mak,Sallen,Wang2012} which opens new possibilities in the emerging field of valleytronics. \cite{Xiao} 

\begin{figure}[t]
    \centering
        \includegraphics[scale=0.12]{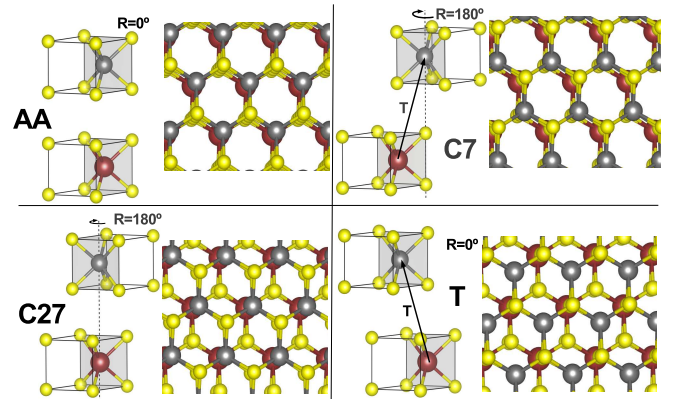}
        \caption{\label{FIG1}(Color online) Schematic views of the MoS$_2-$WS$_2$ heterojunction of different stacking (i.e. C7, C27, AA, T). Each stacking is obtained either by  a monolayers translation T and /or a rotation R with respect to each other. Red, gray, and yellow spheres represent W, Mo, and S atoms respectively.}
\end{figure}

Importantly, other transition metal dichalcoghenides, such as WS$_2$, as well as MoSe$_2$ and WSe$_2$  are expected to have similar properties, \cite{Zhu,Jiang,Ramasubramaniam,Feng_PRB2012} and the first experimental demonstrations of monolayer WS$_2$ have just been reported. \cite{Terrones2012}  All of the above naturally leads us to investigate the electronic properties of transition metal dichalchogenide (TMD) multilayers \cite{Castellanos,Zeng,Wu12}. Here we report our results on the simplest case, a bilayer of MoS$_2$ and WS$_2$, which both have the same crystal structure and very similar lattice constant. In particular we are  interested in how the stacking of different TMD monolayers  (see Fig. \ref{FIG1}) can result in heterostructures with electronic properties different from the  homogeneous TMD monolayers and multilayers, 

Our calculations were performed with the Vienna {\it ab-initio} package ({\small VASP}), \cite{VASP1} based on the  local density-functional  approximation, \cite{KS-PR} plane-wave basis ($E_{\rm cut} = 400$ eV) and non-collinear  projector-augmented waves (PAW) method. \cite{Blochl1,Hoobs} We treat the both the transition metal orbitals $4p$, $5s$, $4d$ together with the Sulphur  orbtials $3s$ and $3p$ as  valence states, and the rest  are considered as  core.  We use the  Perdew-Burke-Ernzerhof's \cite{PBE} version of generalized gradient approximation to describe the exchange correlation density functional. All calculations are carried out using a 1$\times$1 supercell with vacuum thickness not smaller than 17 {\AA}. The $\Gamma$-centered Monkhorst-Pack's \cite{Monkhorst} mesh ($6\times6\times1$) of the \textbf{k}-points was used to sample the BZ.

\begin{figure}[t]
    \centering
    \includegraphics[scale=0.63]{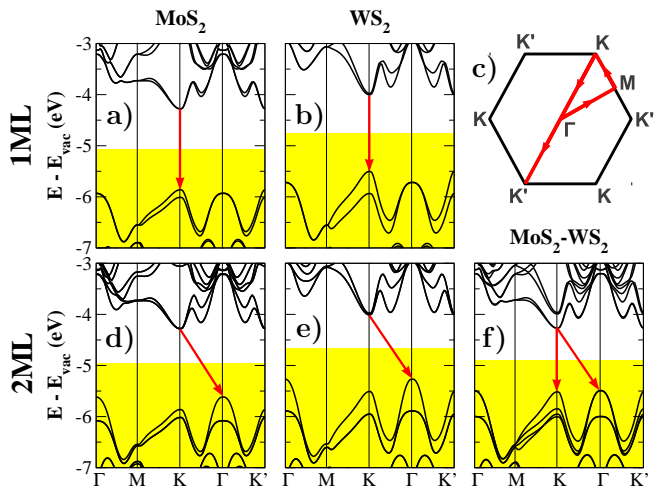}
        \caption{\label{FIG2}(Color online) Band structures of: a) MoS$_2$ monolayer, b)  WS$_2$ monolayer, d) MoS$_2$ bilayer, e) WS$_2$ bilayer, f) MoS$_2-$WS$_2$ heterojunction. The stacking of bilayers is  C7  (see Fig. \ref{FIG1}(b)). c) scheme of the BZ with the line along which the band structures are calculated. E$_{\rm{vac}}$ stands for vacuum energy. The Fermi energy lies at the intersection of white and yellow regions.}
\end{figure}

For reference, we discuss first the  electronic properties of isolated MoS$_2$ and WS$_2$ monolayers (MLs). \cite{Zhu,Jiang,Ramasubramaniam,Feng_PRB2012,Cheiw2012,Kadantseva}
The crystal structure of 2H-MoS$_2$   (2H-WS$_2$) consists of two 2D parallel triangular lattices of S atoms separated by the same lattice of Mo (W)  atoms translated by $1/3$ of the unit-cell diagonal, with  lattice constant $a = 3.19$ {\AA} ($a=3.20${\AA}).\cite{Zhu}  The corresponding  Brillouin zone (BZ) is also hexagonal, with two inequivalent $K$ and $K'$ points (valleys). We show the corresponding energy bands in  Figs. \ref{FIG2}(a) and \ref{FIG2}(b)), which are in agreement with previous work using the same methodology.\cite{Ramasubramaniam,Jiang} Both MLs are direct band semiconductors with a maximum valence band (VB) and minimum conduction band (CB) located in the $K$ and $K'$ valleys. The band-gap values equal 1.58 and 1.50 eV for MoS$_2$ and WS$_2$ respectively. Our calculations also show that,  when referred with respect to the vacuum energy, the band structures of both MLs are shifted  (cf. Figs. \ref{FIG2}(a) with \ref{FIG2}(b)), on account of the different electronegativity of the Mo and W.

Because of the lack of inversion symmetry and a strong spin-orbit coupling (SOC) the valence  and conduction bands  are spin-split at  the $K$ and $K'$ points. The sign of the spin splitting changes from $K$ to $K'$   resulting in the so called strong spin-valley coupling. \cite{Xiao} The splitting is higher in WS$_2$ ML (435 and 27 meV for VB and CB respectively) than MoS$_2$ ML (147 and 3 meV for VB and CB respectively) due to the higher atomic number of W than Mo.

The band-dependence of the spin splittings is accounted for by the atomic orbital composition of the states. Our population analysis reveals that, at the $K$ point, the CB minimum is mostly made by Mo $d_{z^2}$ $(l=2, m=0)$ orbitals, whereas the VB maximum dominant contribution comes from the $d_{xy}$ and $d_{x^2-y^2}$ $(l=2, m=\pm2)$ orbitals. To leading order in the SOC, this should yield a non-zero valley dependent spin-splitting only in the VB, in agreement with the toy model proposed by Xiao \textit{et al.} \cite{Xiao} However,  both VB and CB states at the $K,K'$ points have small contributions coming from the Sulphur $p_x$ and $p_y$ orbitals $(l=1,m=\pm1)$. These are probably behind  the small but finite splitting in the CB.

We now discuss the electronic properties of the bilayers that can be formed stacking the WS$_2$ and MoS$_2$ monolayers. We have verified that the main features of the electronic structure are quite insensitive to the  stacking (see  Fig. \ref{FIG1} for the different stackings),  thereby we focus on the band structure of the  C7 stacking (see Fig. \ref{FIG1}(b)) presented in Fig. \ref{FIG2}(f). This  is the stacking  of bulk  MoS$_2$ and WS$_2$.     Comparison of monolayer and bilayer bands in Fig. 2  indicates that interlayer coupling is not strong.

The electronic structure of Mo-Mo and W-W  bilayers (Figs. \ref{FIG2}(d), (e)) can be rationalized in terms of two concepts:  interlayer coupling of degenerate monolayer states, which splits  most of the monolayer states, and the existence of a symmetry center in the C7  stacking, which prevents spin splittings. The interlayer splitting is significantly stronger for the valence band at the $\Gamma$ point than for the VB and CB at K points. As a result, the highest VB state moves to the $\Gamma$ point for the W-W and Mo-Mo bilayers, which become indirect gap systems. 

In the case of the  Mo-W heterojunction  the interlayer coupling competes with the energy difference of the monolayer states,  shown in Fig. 2 (a),(b).   As a result, the   VB at the $\Gamma$ point is almost degenerate ($\Delta E_{\rm VB}=27$ meV) with the top of the VB at the  K and K'  points. Consequently, a significant population of photoexcited holes will be available at the K and K' points, and photoluminescence will be not quenched.  In this sense, the MoS$_2-$WS$_2$ heterojunction -- unlike the homogeneous bilayers -- will be optically active. In addition, the Mo-W bilayer does not have inversion symmetry, so that spin splittings at the K and K' points occur like in the monolayers.

\begin{figure*}[!t]
    \centering
        \includegraphics[scale=0.32]{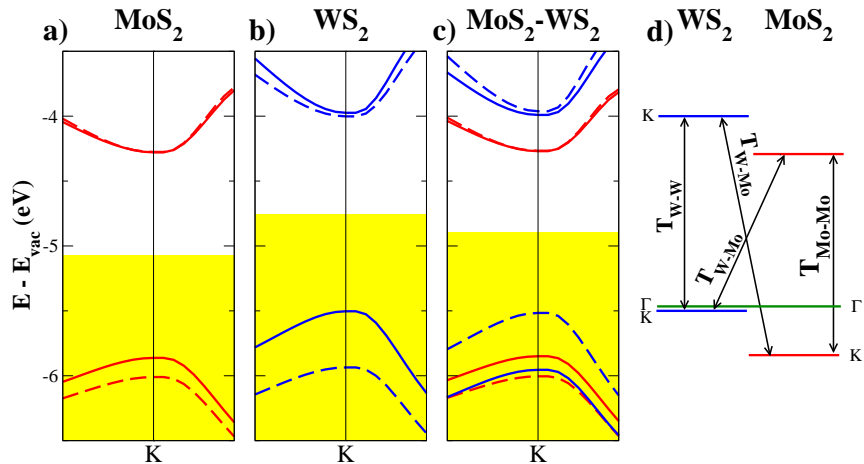}
        \caption{\label{FIG3}(Color online) Zoom of  band structures at the $K$ point for : a) MoS$_2$ monolayer, b)  WS$_2$ monolayer, c) MoS$_2-$WS$_2$ bilayer, d) scheme of possible optical transitions in the MoS$_2-$WS$_2$ bilayer. Blue (red) lines describe the bands of the states localized on the W (Mo) atoms. Bold (dashed) lines describe the states of spin up (down).}
\end{figure*}

A summary of the electronic states  in the neighborhood of the $K$ point,  relevant for the inter-band optical experiments, both for MoS$_2$, WS$_2$ and the Mo-W bilayer is shown in Fig. \ref{FIG3}. It is apparent that  interlayer coupling at this point is negligible and the bilayer bands are nothing but a superposition of the monolayer states.  As a result, the top of the VB is in the W layer and the bottom of the CB name lies in the Mo layer, forming a type II structure. \cite{Cardona-Yu} In addition the Mo-W bilayer  gap is  1.2eV,  0.3eV smaller than the gap of the monolayers. In contrast, the top of the VB at the $\Gamma$ point is delocalized in both planes.   The resulting scheme of levels is shown in Fig. 3(d).   We expect that intra-layer transitions have a stronger quantum yield, on account of their larger electron-hole overlap,\cite{Cardona-Yu} but relaxation to the lower energy spatially separated electron-hole pair is expected. 

We finally comment on the limitations of our methodology.  First,  PBE is known to underestimate the band gap. The use of either  non-local functionals \cite{Ramasubramaniam} and/or GW approximation \cite{Jiang,Cheiw2012}  yields a better agreement with the experiments.   Second, excitonic effects, not included in the band structure calculation, are known to produce a large shift in the absorption threshold and photoluminescence peaks. \cite{ Ramasubramaniam} In spite of this, we expect that the band-gap reduction and the segregation of electrons and holes in different atomic planes will be confirmed by experimental work and/or  more sophisticated  methodologies. 

In summary, we have studied  the electronic properties of the MoS$_2-$WS$_2$ system as an example of transition metal dichalcogenide two-dimensional heterostructure. We find that, in contrast to the Mo-Mo and W-W bilayers, the band-gap is direct. Additionally we find that the lowest energy electron and highest energy hole  states in the optically active $K$ point are localized on different monolayers. In this sense, the Mo-W bilayer forms a type II heterostructure. 
The combination of band gap engineering in heterojunctions found here, together with the reported electrical control of electronic and optical properties in these systems\cite{Radisavljevic,Zeng,Wu12,Bolotin},  hold the promise of a bright future for optospintronics in  two dimensional transition metal dichalcogenides.

We acknowledge J. W. Gonz\'alez and F. Delgado for  fruitful discussions. This work has been financially supported by MEC-Spain (Grant Nos. FIS2010-21883-C02-01, FIS2009-08744,  and CONSOLIDER CSD2007-0010) and Generalitat Valenciana, grant Prometeo 2012-11.


\begin{references}

\bibitem{Alferov} Z. I. Alferov, Review of Modern Physics {\bf 73}, 767 (2001)

\bibitem{Cardona-Yu}  P. Y. Yu and M. Cardona, {\em Fundamentals of Semiconductors}, Springer, New York (1996)

\bibitem{VK} K. Von Klitzing,  Review of Modern Physics {\bf 58}, 519 (1986)

\bibitem{Tsui} D. C. Tsui, H. L. Stormer, and A. C. Gossard, Phys. Rev. Lett.  {\bf 48}, 1559 (1982).

%\bibitem{Laughlin} R.B. Laughlin, PRL {\bf 50}, 1395 (1983).

\bibitem{Butov.PRL} L. V. Butov, A. L. Ivanov, A. Imamoglu, P. B. Littlewood, A. A. Shashkin, V. T. Dolgopolov, K. L. Campman,
and A.C. Gossard, PRL {\bf 86}, 5608 (2001).


%Title: High-temperature ultrafast polariton parametric amplification in semiconductor microcavities
\bibitem{Saba}   M. Saba, C. Ciuti, J. Bloch {\em et al.}. Nature {\bf 214}, 731 (2001)

%Title: Condensation of semiconductor microcavity exciton polaritons
\bibitem{Deng} H.  Deng, G.  Weihs, C. Santori, C {\em  et al.}, Science {\bf 298}, 199 (2002)

\bibitem{Konig} M. K\~onig, S. Wiedmann, C. Brune, A. Roth, H. Buhmann, L. W. Molenkamp, X. L. Qi, and S. C. Zhang, Science {\bf 318}, 766 (2007). 

%BEC

%\bibitem{Butov.Nature} L.V. Butov, C.W. Lai, A.L. Ivanov, A.C. Gossard, and D.S. Chemla, Nature {\bf 417}, 47 (2002).

\bibitem{PNAS} K. S. Novoselov, D. Jiang, F. Schedin, T. J. Booth, V. V. Khotkevich, S. V. Morozov, and A. K. Geim, Proc. Nat. Ac. Sci.  {\bf 102}, 10451 (2005).

%Geim Nature 2005
\bibitem{Novoselov05} K. S. Novoselov {\em et al.}, Nature {\bf 438}, 197 (2005).

%Kim Nature 2005
\bibitem{Kim05} Y. Zhang {\em et al.}, Nature {\bf 438}, 201 (2005).

\bibitem{Novoselov2012} K. S. Novoselov and A.H. Castro Neto, Phys. Scr. {\bf T146}, 014006 (2012).

%Title: Unconventional quantum Hall effect and Berry's phase of 2 pi in bilayer graphene
\bibitem{Novoselov06} K. S.  Novoselov, E. McCann, S. V: Morozov, SV; {\em et al.} Nature Physics  {\bf 2} 177 (2006)

% Boron nitride substrates for high-quality graphene electronics.
\bibitem{Dean} C. R. Dean, {\em et al.} Nature Nanotech. {\bf 5}, 722 (2010).



\bibitem{Remskar} M. Remskar, A. Mrzel, Z. Skraba, A. Jesih, M. Ceh, J. Demšar, P. Stadelmann, F. Lévy, and D. Mihailovic, Science {\bf 292}, 479 (2001).

\bibitem{Enyashin} A. N. Enyashin, L. Yadgarov, L. Houben, I. Popov, M. Weidenbach, R. Tenne, M. Bar-Sadan, and G. Seifert, J. Phys. Chem. C {\bf 115}, 24586 (2011).

\bibitem{Deepak} F.L. Deepak, H.C., S. Cohen, Y. Feldman, R. Popovitz-Biro, D. Azulay, O. Millo, and R. Tenne, J. Am. Chem. Soc. {\bf 129}, 12549 (2007)



\bibitem{Helveg-PRL2000} S. Helveg {\em et al.}, Phys. Rev. Lett. {\bf 84}, 951  (2000)

%Structure of single-molecular-layer MoS2
\bibitem{Yang91} D. Yang, S. Jim\'enez-Sandoval, W. M. R. Divigalpitiya, J. C. Irwin, R. F: Drid,  Phys. Rev. B.{\bf 43},  12053 (1991)

\bibitem{Splendiani} A. Splendiani, L. Sun, Y. Zhang, T. Li, J. Kim, C.-Y. Chim, G. Galli, and F. Wang, Nano Lett. {\bf 10}, 1271 (2010).

\bibitem{Mak_PRL2010} K. F. Ma, C. Lee, J. Hone. J. Shan, T. F. Heinz, Phys. Rev. Lett. {\bf 105}, 136805 (2010)

\bibitem{Radisavljevic} 
%Single-layer MoS2 transistors
B. Radisavljevic, A. Radenovic, J. Brivio, V. Giacometti, and A. Kis, Nature Nanotechnology {\bf 6}, 147 (2011).

%%% VAlleytronics
\bibitem{Xiao} D. Xiao, G.B. Liu, W. Feng, X. Xu, and W. Yao, Phys. Rev. Lett. {\bf 108}, 196802 (2012).

\bibitem{Cao} T. Cao \textit{et al.}, Nature Communications {\bf 3}, 887 (2012).

\bibitem{ZengNature} H. Zeng, J. Dai, W. Yao, D. Xiao, and X. Cui, Nat. Nano.  {\bf 3}, 490 (2012).

\bibitem{Mak} K.F. Mak, K. He,J. Shan, and T.F. Heinz, Nature Nano. {\bf 7}, 494 (2012).


%Robust optical emission polarization in MoS2 monolayers through selective valley excitation G. Sallen \textit{et al.}, 
\bibitem{Sallen}  G. Sallen, L. Bouet, X. Marie, G. Wang, C. R. Zhu, W. P. Han, Y. Lu, P. H. Tan, T. Amand, B. L. Liu, and B. Urbaszek, Phys. Rev. B {\bf 86}, 081301(R) (2012).

\bibitem{Wang2012}  
%
%REVIEW: Electronics and optoelectronics of two-dimensional transition metal dichalcogenides
Q.  Wang, K. Kalantar-Zadeh, A.  Kis {\em et al.}
Nature Nano. {\bf 7}, 699 (2012)


%MoSe, MoSe2, WS2, SSe2
\bibitem{Zhu} Z.Y. Zhu, Y.C. Cheng, and U. Schwingenschl\"{o}gl, Phys. Rev. B {\bf 84}, 153402 (2011).

%GW calks of TMD
\bibitem{Jiang} H. Jiang, J. Phys. Chem. C {\bf 116}, 7664 (2012).

\bibitem{Ramasubramaniam} A. Ramasubramaniam, Phys. Rev. B {\bf 85}, 115409 (2012).

%Intrinisc Spin Hall effect in monolayers of group VI dichalcogenides: a first principles study
\bibitem{Feng_PRB2012} W. Feng, Y. Yao, W. Zhu, J. Zhou, W. Yao, D. Xiao, Phys. Rev. B {\bf 86}, 165108 (2012)

\bibitem{Terrones2012} H. Guti\'errez \textit{et al.},  arXiv.org 1208.1325 (2012).

%MULTILAYER PAPERS

\bibitem{Castellanos} A. Castellanos-Gomez E. Cappelluti, R. Roldan, N. Agrait, F. Guinea, G. Ruibio-Bollinger, 
Adv. Mater.. doi: 10.1002/adma.201203731

\bibitem{Zeng}H. Zeng \textit{et al.}, 	arXiv 1208.5864 (2012).

\bibitem{Wu12} S. Wu \textit{et al.},  arXiv 1208.6069 (2012)
% Electrical tuning of Valley Magnetic Moment via Symmetry Control

%\bibitem{Lebegue} 
%% MoS2 and NbSe2
%S. Leb\'{e}gue and O. Eriksson, Phys. Rev. B {\bf 79}, 115409 (2009).

\bibitem{VASP1} %G. Kresse and J. Hafner, Phys. Rev. B {\bf 47}, 558 (1993).\\
                G. Kresse and J. Furthm\"uller,  Computat. Mater. Sci. {\bf 6}, 15 (1996).
                G. Kresse and J. Furthm\"uller, Phy. Rev. B {\bf 54}, 11169 (1996).
\bibitem{KS-PR}  W. Kohn and L.J. Sham, Phys. Rev. {\bf 140}, A1133 (1965).
\bibitem{Blochl1} P. E. Bl\"ochl, Phys. Rev. B {\bf 50}, 17953 (1994).
\bibitem{Hoobs} D. Hobbs, G. Kresse and J. Hafner, Phys. Rev. B {\bf 62}, 11556 (2000).
\bibitem{PBE} J. P. Perdew, K. Burke, and M. Ernzerhof, Phys. Rev. Lett. {\bf 77}, 3865 (1996).
\bibitem{Monkhorst} H. J. Monkhorst and J. D. Pack, Phys. Rev. B {\bf 13}, 5188 (1976).

%\bibitem{VASP2}                G. Kresse and D. Joubert, Phys. Rev. B {\bf 59}, 1758 (1999).\\

\bibitem{Cheiw2012} T. Cheiwchanchamnagij, W. R. L. Lambrecth Phys. Rev. B {\bf 85}, 205302 (2012).
\bibitem{Kadantseva}  E.S. Kadantseva, and P. Hawrylak, Solid State Commun. {\bf 152}, 909 (2012).

\bibitem{Bolotin} A. K. M. Newaz {\em et al}, Arxiv 1211.0341


\end{references}
\end{document}